\documentclass[sigconf,10pt,nonacm]{acmart}

\AtBeginDocument{%
  }

\usepackage{multirow}
\usepackage{tikz}  
\usepackage{amsmath} 
\usepackage{xcolor}
\usepackage{enumitem}  
\usepackage{longtable}  
\usepackage{multicol}  
\usepackage{fancyvrb} 
\usepackage[most]{tcolorbox}  
\newcommand{\coloredcircle}[2]{  
  \tikz[baseline=(char.base)]{  
    \node[shape=circle, fill=#1, inner sep=0.8pt, text=white, font=\bfseries] (char) {#2};}} 
\newcommand{\ourmethod}{\textsc{ServiceOdyssey}~}

\definecolor{color1}{HTML}{8EC4E8}
\definecolor{color2}{HTML}{FEA617}
\definecolor{color3}{HTML}{7F8B40}

\begin{document}

\title{Enabling Autonomic Microservice Management through Self-Learning Agents}

\settopmatter{authorsperrow=4}
\author{Fenglin Yu\footnotemark[1]}
\affiliation{%
  \institution{Wuhan University}
  \country{}
}

\author{Fangkai Yang}
\affiliation{%
  \institution{Microsoft}
  \country{}
}

\author{Xiaoting Qin}
\affiliation{%
  \institution{Microsoft}
  \country{}
}

\author{Zhiyang Zhang\footnotemark[1]}
\affiliation{%
  \institution{Nanjing University}
  \country{}
}

\author{Jue Zhang\footnotemark[2]}
\affiliation{%
  \institution{Microsoft}
  \country{}
}

\author{Qingwei Lin}
\affiliation{%
  \institution{Microsoft}
  \country{}
}

\author{Hongyu Zhang}
\affiliation{%
  \institution{Chongqing University}
  \country{}
}

\author{Yingnong Dang}
\affiliation{%
  \institution{Microsoft}
  \country{}
}

\author{Saravan Rajmohan}
\affiliation{%
  \institution{Microsoft}
  \country{}
}

\author{Dongmei Zhang}
\affiliation{%
  \institution{Microsoft}
  \country{}
}

\author{Qi Zhang}
\affiliation{%
  \institution{Microsoft}
  \country{}
}

\pagestyle{plain}
\begin{abstract}

The increasing complexity of modern software systems necessitates robust autonomic self-management capabilities. While Large Language Models (LLMs) demonstrate potential in this domain, they often face challenges in adapting their general knowledge to specific service contexts. To address this limitation, we propose \textbf{\textsc{ServiceOdyssey}}, a self-learning agent system that autonomously manages microservices without requiring prior knowledge of service-specific configurations. By leveraging curriculum learning principles and iterative exploration, \ourmethod progressively develops a deep understanding of operational environments, reducing dependence on human input or static documentation. A prototype built with the Sock Shop microservice demonstrates the potential of this approach for autonomic microservice management.
\end{abstract}




\maketitle
\renewcommand{\thefootnote}{\fnsymbol{footnote}}
\footnotetext[1]{Work is done during an internship at Microsoft.}
\footnotetext[2]{Corresponding author (juezhang@microsoft.com).}

\thispagestyle{plain}

\section{Introduction}

Modern software systems are becoming increasingly distributed and large-scale, posing significant management challenges. This growing complexity underscores the importance of autonomic self-management capabilities~\cite{Kephart2003TheVO}. Recent advancements in Large Language Models (LLMs), known for their exceptional contextual understanding and adaptive decision-making capabilities~\cite{zhang2024survey}, have inspired new approaches of developing LLM-based agents for autonomic software service management~\cite{jin2024llms}.

Despite their potential, leveraging LLMs for autonomic microservice management presents challenges. A key issue is the knowledge gap between the general knowledge embedded in LLMs and the specific knowledge required for managing particular services. For example, LLM-based agents lack prior awareness of service-specific configurations, such as monitoring setups using Prometheus~\cite{prometheus_website}. 
A straightforward solution involves manually incorporating these configurations into the agents' prompts, as implemented in~\cite{ACV24Zhang}. Alternatively, Retrieval-Augmented Generation (RAG)~\cite{lewis2020retrieval} can be used  to extract relevant details from service documentation.
However, manual integration is labor-intensive, and documentation-based approaches depend heavily on the quality and availability of the underlying resources. Furthermore, the dynamic nature of software systems demands agents capable of continuously updating their knowledge.

To overcome these limitations, we present a self-learning agent system tailored to microservice management, a prevalent architecture for orchestrating complex distributed cloud services. Unlike the above approaches that mainly rely on human input or service documentation, our approach enables agents to autonomously and continuously acquire knowledge through interaction with the operational environment. The system starts without specific knowledge of the managed service, reducing the reliance on human-provided details, and builds understanding progressively through self-directed exploration tasks. These tasks are structured using curriculum learning principles to facilitate systematic knowledge acquisition. Drawing an analogy to a voyager exploring unknown territories, we named this self-learning agent system \textbf{\ourmethod} to reflect its systematic exploration and comprehension of unfamiliar microservices. 

The subsequent sections outline the design of \ourmethod, followed by a prototype implementation utilizing the Sock Shop microservice demo project~\cite{sockshop}.

\section{Design of the \ourmethod System}
\begin{figure}[t]
  \centering
  \includegraphics[width=0.95\linewidth]{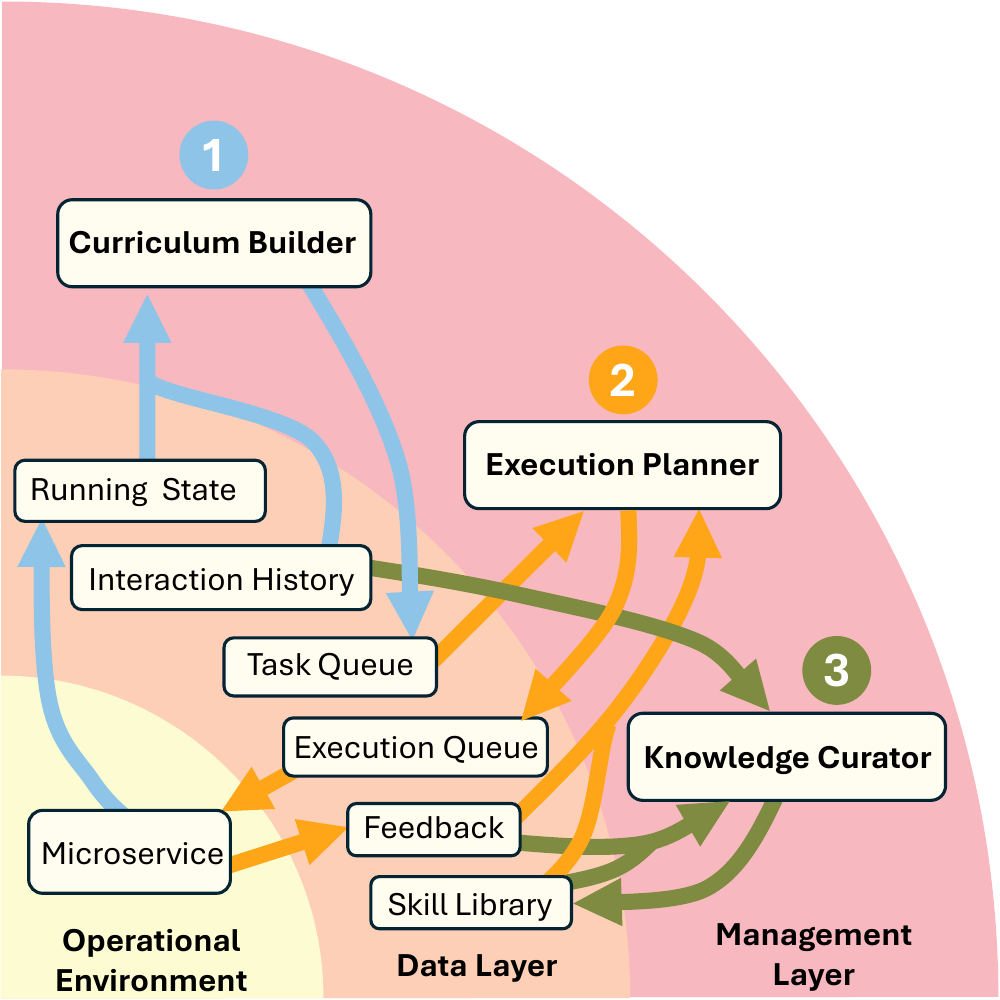}
  \caption{An overview of the \ourmethod system.}
  \label{fig:overview}
  \vspace{-5mm}
\end{figure}

The \ourmethod system architecture, shown in Figure~\ref{fig:overview}, is built on a microservice system augmented by a data layer and a management layer. This design enables efficient self-learning and task execution within microservices. The \textbf{data layer} serves as a repository for critical data, including system running state and interaction history. This information is passed to the management layer in self-learning processes. Additionally, the data layer stores data during self-learning such as tasks and execution plans, which are subsequently executed in the operational environment. The data layer works as an intermediate layer, bridging the microservice system and the LLM-enhanced management layer. The \textbf{management layer} consists of three key LLM-enhanced modules: \coloredcircle{color1}{1}~\textbf{Curriculum Builder}~(CB) module progressively generates tasks, aiming to explore and learn the microservice system. \coloredcircle{color2}{2}~\textbf{Execution Planner}~(EP) module is responsible for generating executable plans and actions by translating tasks generated by 
CB into actionable strategies. These strategies are executed within the microservice environment, with real-time feedback and interaction history being collected. The \coloredcircle{color3}{3}
~\textbf{Knowledge Curator}~(KC) module consolidates the feedback and interaction history from task execution into a comprehensive skill library. This growing collection of skills can be utilized by 
EP to tackle increasingly complex management tasks in the microservice system.

\subsection{Curriculum Task Generation}

When facing a new microservice system, humans often apply their existing knowledge to explore and understand the unfamiliar environment. This process, known as \textit{discovery learning} in educational theory~\cite{bruner1961act}, involves actively using their prior knowledge to independently explore new concepts. Inspired by this approach, we designed 
~\textbf{Curriculum Builder}~(CB) to generate tasks that facilitate the exploration and learning of new microservice systems.

\noindent\textbf{Task generation principles.}
We design directives to prompt LLMs to generate diverse tasks with principles, helping explore and discover the system. The generated tasks should reflect the progressive learning mechanism, building upon previous experienced tasks:


\begin{enumerate}[label=({\arabic*}), left=0pt]  
    \item \textit{from easy to hard}:
    Following the curriculum learning methodology~\cite{bengio2009curriculum,wang2021survey}, tasks are generated in a sequence that progresses from easy to hard. The CB reviews previously completed tasks to generate new ones, building on successful experiences by adding incremental challenges. Successful tasks guide the creation of similar, yet more challenging tasks. In the meanwhile, failed tasks prompt the generation of easier or alternative tasks that could potentially be solved.   
    \item \textit{from observation to action}:
    Following the DevOps practices~\cite{jabbari2016devops,chen2018microservices}, the transition from observation to action is a fundamental aspect of managing microservice systems. Initially, tasks focus on observing the system to gather insights into its performance and configuration, using monitoring tools and metrics. These observation tasks are safe and do not alter the system state. Once a comprehensive understanding is achieved, the focus shifts to action tasks, where changes are implemented to optimize the system based on observations. These actions may include configuration tuning, resource scaling, and service refactoring to address issues like latency reduction and resource efficiency. This curriculum task building, central to DevOps, ensures continuous understanding and management of the microservice system. 
\end{enumerate}

\noindent\textbf{Context for task generation.} 
To better prompt LLMs to generate curriculum tasks, the input consists of several parts: 
\begin{enumerate}[label=({\arabic*}), left=0pt]  
    \item \textit{System running state}: This includes contextual information such as traffic load, system health status, etc. The dynamic nature of the running state allows CB to create tasks that are highly relevant to the current status of the microservice. For instance, a normal running state might prompt monitoring and logging tasks, while an anomalous state could lead to tasks focused on diagnosis and mitigation.      
    \item \textit{Interaction history}: 
    This records the details and outcomes of previous interactions between agents and the managed microservice, reflecting the current management capabilities and self-learning progress. Including it in CB would provide detailed context to generate new tasks following the generation principles. For example, following the ``from observation to action'' principle, given the history containing the task of acquiring the CPU usage, the new tasks from CB could be reducing the CPU usage, which is built upon the successfully learned skills.     
    \item \textit{Other information}: Although not shown in Figure~\ref{fig:overview}, other resources (if they are accessible), such as the microservice documentation, codebase, and available tools, can further help CB generate more relevant tasks. 
\end{enumerate}


\subsection{Solution Refinement with Feedback}\label{sec:solutionrefinement}
Once the tasks are generated from 
~CB and put into the task queue, 
~\textbf{Execution Planner}~(EP) aims to generate executable solutions leveraging previously learned skills and feedback. However, LLMs have challenges to ensure solutions to be executable in specific microservice environment~\cite{wang2023survey}. To address this challenge, we design a solution refinement mechanism that incorporates collective feedback to iteratively reflect and refine the generated solutions to make them executable and valid in the microservice system. High-quality solutions are crucial for 
~KC to create valid and useful skills.

\begin{figure}[t]
  \centering
  \includegraphics[width=\linewidth]{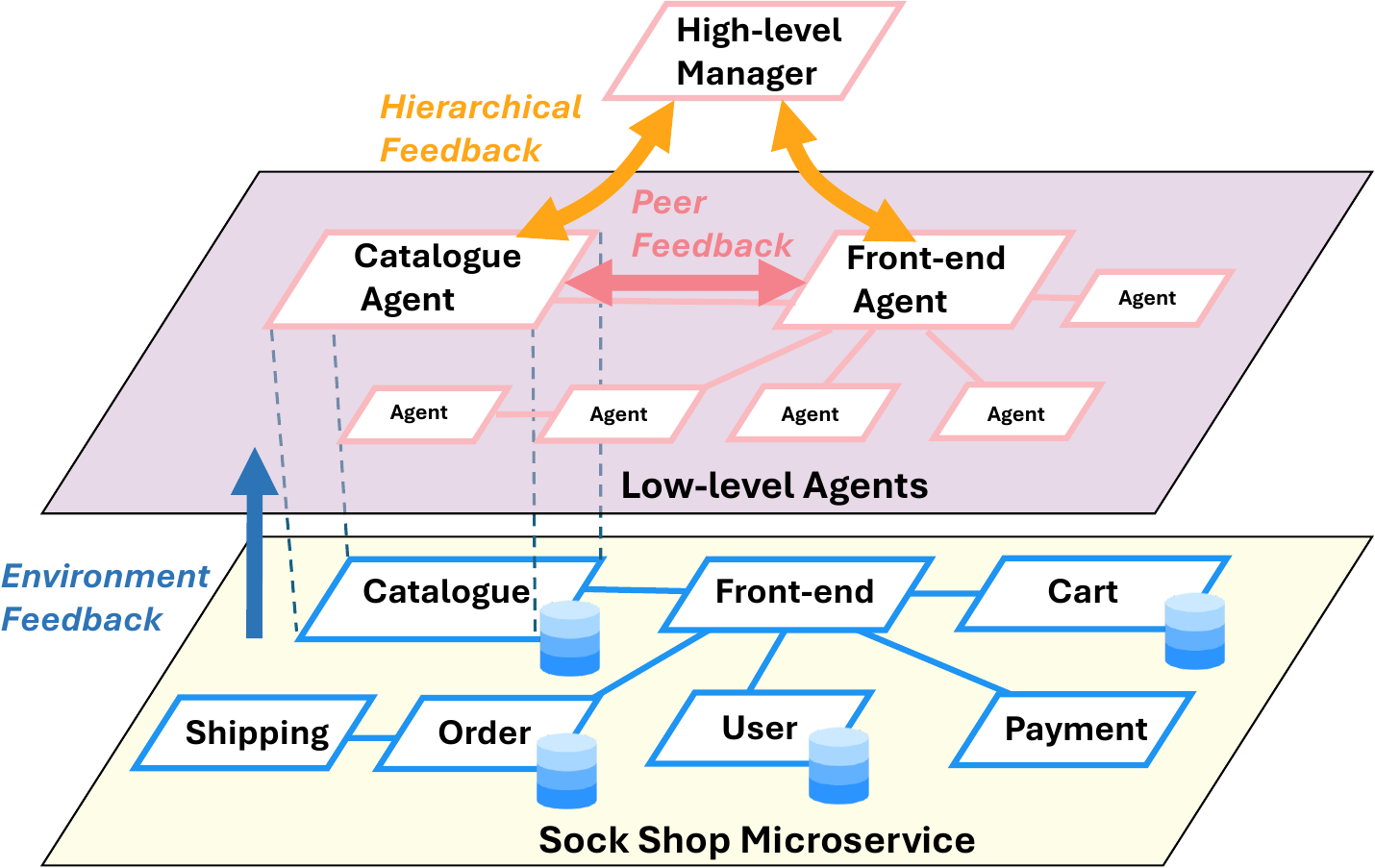}
  \caption{Agentic microservice system for Sock Shop. }
  \label{fig:selflearn}
  \vspace{-4mm}
\end{figure}

\noindent\textbf{Agentic microservice system.}
We first introduce the environment where the solution is executed. For illustration, we adopt the Sock Shop microservice~\cite{sockshop}, which simulates an e-commerce platform for selling socks, as a representative microservice environment. To enable LLMs to provide feedback in the solution refinement, the Sock Shop microservice system turns into an agentic microservice system with each system component managed by an LLM-enhanced agent~\cite{ACV24Zhang} (as shown in Figure~\ref{fig:selflearn}). For example, the \textit{Catalogue} component turns into the \textit{Catalogue} agent, supporting natural language-based management and communication. These agents form the low-level agents layer in Figure~\ref{fig:selflearn}. Above it, an LLM-based high-level manager is designed as an entry point to decompose tasks into subtasks and coordinate low-level agents~\cite{khot2022decomposed,prasad2023adapt}. 

\noindent\textbf{Solution initialization.} 
The solution generation process begins with the high-level manager analyzing the task description and retrieving pertinent skills from the skill library~\cite{gao2023retrieval} for task decomposition. If no suitable skills are available, the high-level manager leverages the inherent knowledge of LLMs. Subtasks are then allocated to low-level agents in order to generate executable actions to address these subtasks. Once all subtasks are completed, their solutions are ensembled to form the final solution. In practice, it is challenging to generate an executable and valid solution in one attempt, and it often requires reflection and refinement from feedback to make solutions executable and valid.   

\noindent\textbf{Reflection and refinement from feedback.}
There are typical three types of feedbacks, each type providing specific refinement suggestions: 

\begin{enumerate}[label=({\arabic*}), left=0pt]  
    \item \textit{Environment feedback}: This feedback contains syntax and execution errors from the microservice system, providing direct and reliable feedback to improve solution quality and fix bugs. LLMs can learn from system error messages to refine solutions~\cite{fan2023automated,xia2023automated}. Such environment feedback reveals the microservice properties such as configurations, enabling syntax corrections and solution adaptations to suit the microservice environment.
    \item \textit{Peer feedback}: Task completion often involves multi-agent collaboration~\cite{guo2024large,xu2024magic}. Peer feedback occurs from the collaboration among ``peer'' agents at the same hierarchical level (or neighboring agents in the decentralized setting). The execution of a subtask can depend on the execution results from the former agent. An agent might self-verify its solution to resolve an assigned subtask limited to the knowledge scope of itself. However, the feedback from downstream agent provides extra requirements to refine the solution. One typical example is the format of the execution results from an agent does not meet the input requirement of the downstream agent, and the format requirement feedback helps modify the solution to ensure the ensembled solution from involved agents to be valid and executable.
    \item \textit{Hierarchical feedback}: The high-level manager oversees the entire subtask completion process and dynamically alters the task decomposition and allocation based on the task execution process. In practice, the initial task decomposition is not always optimal~\cite{huang2024understanding}, the high-level manager do re-decomposition and re-planning in an interleaving manner and modifies the subtask allocation and content. Additionally, the hierarchical feedback provides global and indirect information, which the peer feedback cannot support. For example, a task of reducing the overall system's latency requires the collaboration of low-level agents to reduce latency. Different low-level agents have diverse capabilities in latency reduction, and the hierarchical feedback adjusts the latency reduction objective for each agent based on its capability. This hierarchical feedback refines the solution to be more cohesive and optimal.
\end{enumerate}  

\subsection{Knowledge Creation and Validation}
The aim of 
~\textbf{Knowledge Curator}~(KC) is to consolidate the knowledge from the successfully executed tasks with related solutions into a skill library. The skill library represents the knowledge of the microservice system after self-learning and is valuable for self-management of the system. We define three skill schemas: \textit{Command}, \textit{Reflection}, and \textit{Configuration} in the Sock Shop microservice for 
~KC to better extract skills. Below shows skill examples with each type: 

{\scriptsize
\begin{tcolorbox}[colback=gray!10, colframe=gray!80, title= Skill Example]  
  \textbf{Command:} \texttt{query\_prometheus(promQL='count by (\_\_name\_\_)({job=\\"sock-shop/catalogue"})', duration='2m', step='1m')} \\  
  \textit{Description:} Can be used for retrieving the list of metrics currently being collected by Prometheus for \textit{Catalogue}.\\
  \textbf{Reflection:} Continuous empty results in Prometheus queries could indicate misconfigured metric collection or incorrect label usage.\\
  \textbf{Configuration:} The \textit{Catalogue} component in \texttt{sock-shop} namespace uses the label \texttt{name=catalogue}.  
\end{tcolorbox} 
}


The above skill schema is generally applicable in other microservice systems. The \textit{Command} type focuses on executable solutions, the \textit{Reflection} type focuses on thoughts that lead to successful task completion, and the \textit{Configuration} reflects the property of the microservice system.


After the skill creation, there is a validation process. The hallucination problem~\cite{huang2023survey} is a big challenging in LLM-based generation, and the created skills face the similar problem in code validity~\cite{apifirst}. The validation process execute the skill in microservice system to check if there is any system error. LLMs are then use to verify if the execution results matches the expected output of the skill.


\begin{figure}[t]
  \centering
  \includegraphics[width=\linewidth]{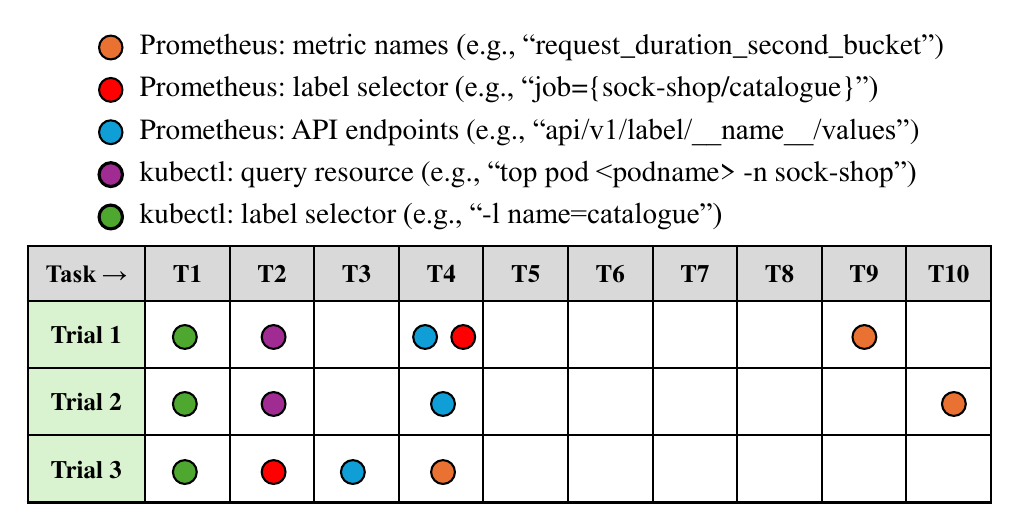}
  \caption{Knowledge curation during self-learning.}
  \label{fig:learned_skills}
  \vspace{-3mm}
  \end{figure}

\begin{figure}[t]
  \centering
  \includegraphics[width=\linewidth]{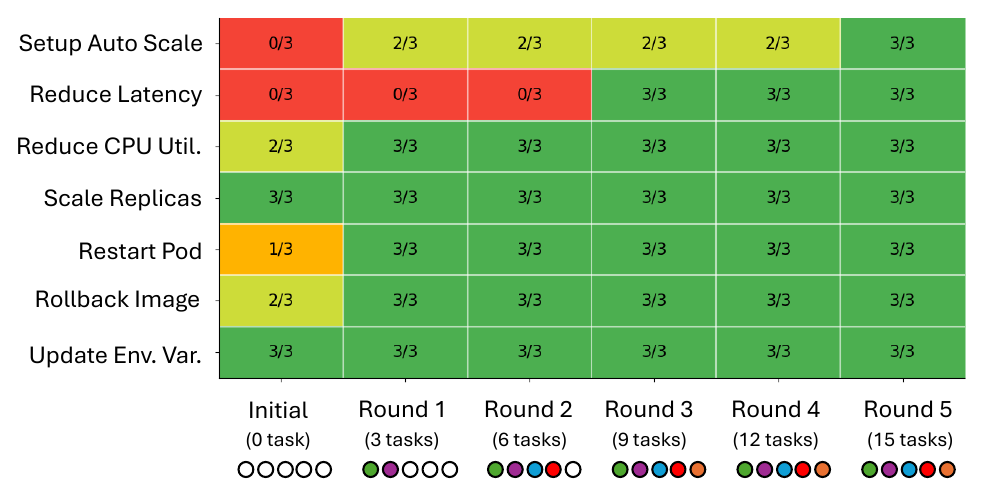}
  \caption{Performance evaluation in Trial 1 after each exploration round. The Y-axis lists evaluation tasks; each cell’s “x/y” indicates the number of successful trials out of the total.}
  \label{fig:heatmap}
  \vspace{-3mm}
\end{figure}

\subsection{\ourmethod in Action}
The three modules work together to self-learn the real-world microservice systems. This workflow supports diverse deployment setups. Ideally, if a canary environment is available, \ourmethod runs in full functionality and adjusts the system state freely. In settings without such a sandbox, it restricts itself to safer observation tasks (e.g., monitoring) while still learning system structures and component properties. Additionally, if manual diagnosis and mitigation logs exist, KC can convert these into new skills, reducing future human intervention whenever similar issues recur.


\section{Prototype Implementation}


We implemented a prototype of \ourmethod in Python, applying it to the Sock Shop microservice. Building on the codebase of~\cite{ACV24Zhang}, we incorporated additional components, including the Curriculum Builder, Knowledge Curator modules, and data-layer elements like the skill library. Hierarchical agent architecture is also adopted for the Execution Planner, while removing Sock Shop-specific instructions in prompts for broader microservice applicability (e.g., the label selector ``\textit{-l name=Catalogue}''). Given the Execution Planner's main role in managing routine operations, we employ \texttt{GPT-4o}~\cite{openai2024gpt4o} as its base LLM to prioritize its low-latency requirement. Meanwhile, the \texttt{o1} model~\cite{openai2024gpto1}, with its advanced reasoning capabilities, is integrated into Curriculum Builder and Knowledge Curator, as their self-learning processes are often confined to non-peak service periods. The other system deployment replicated the setup described in~\cite{ACV24Zhang}. 

In this preliminary study, we evaluate how agents acquire service-specific knowledge through observation-driven exploration tasks from the ground up. As in~\cite{ACV24Zhang}, we utilized a three-agent system comprising a high-level manager overseeing two agents for \textit{Catalogue} and \textit{Front-end}, respectively. Three experimental trials were conducted, each starting with the Curriculum Builder assigning exploration tasks to the high-level manager, who then coordinated their execution by the lower-level agents. Tasks were generated in rounds, with three tasks per round, and concluded after five rounds. A prompt except used for task generation is shown below:

{\scriptsize
\begin{tcolorbox}[colback=blue!10, colframe=blue!70, title= Prompt Excerpt for Task Generation]  
    For the tasks that do not change the system state, the goal is to help the agent familiarize with the service component and its basic metrics. You should generate them in the following order: \\
    \textbf{1.} Check the basic information of the component, such as the version, status, and configuration. \\
    \textbf{2.} Check the basic resource usage of the component, such as CPU and memory usage. They should be obtainable with built-in ``kubectl'' commands.\\
    \textbf{3.} Find out what middlewares are used in the component, such as Redis, Kafka, Prometheus. \\
    \textbf{4.} If Prometheus is used in the component, try to get familiar with its basic usage, such as how to query metrics related to latency and error rates. If the agent is struggling with metric collection via Prometheus, you should further break down the task into smaller steps, such as identifying current Prometheus settings, what the metrics are, and how to query them.  
\end{tcolorbox} 
}

In Figure~\ref{fig:learned_skills}, we depict the knowledge curation process observed during three self-learning trials. Five key knowledge points, marked as colored dots, summarize the accumulated knowledge. Two points pertain to correctly constructing “\textit{kubectl}” commands with service-specific settings, while three focus on Prometheus configuration. The learning process consistently begins with “\textit{kubectl}”-related tasks before transitioning to Prometheus settings, mirroring the task generation order shown in the above prompt excerpt. Despite this consistency, the trials exhibit variation due to the probabilistic behavior of LLMs and dynamic environments. The first two trials show similar trajectories, with rapid acquisition of initial knowledge points followed by challenges in learning Prometheus-related metric queries (i.e., the one in orange dot), particularly latency-related tasks. In contrast, the third trial demonstrates faster progress in mastering Prometheus-related skills.\footnote{The absence of knowledge on querying resources via \textit{kubectl} is attributed to the agent opting for Prometheus to fulfill the corresponding tasks.}

To evaluate the effectiveness of learned knowledge, we evaluated the system at the end of each exploration round in Trial 1 using the operational tasks outlined in~\cite{ACV24Zhang}. Tasks that altered system states were chosen to ensure no overlap with those used in self-exploration. Each evaluation task was repeated three times, with detailed results presented in Figure~\ref{fig:heatmap}. The findings demonstrate a clear correlation between accumulated service knowledge and task performance, with notable improvements in the early rounds. The system is also cost-effective, with each trial costing less than \$10 in terms of LLM API usage and completing in under 30 minutes. 

\section{Related Work}
\noindent\textbf{Self-Learning Agent}. A pivotal advancement in agent design lies in the integration of memory modules~\cite{survey_memory24Zhang}, which enable agents to retain and build upon past experiences. This capability enables agents to develop more robust and adaptive behaviors over time. In the context of LLM agents, exploration-based self-learning has been explored cross various domains, including gaming~\cite{voyager, agentRLgame24Xu}, web automation~\cite{WebAgent23Gur}, computer control~\cite{agent_computeruse24Tan}, smartphone usage~\cite{agent_phoneuse23Zhang}, social interactions~\cite{social_selfevolve24Guan}, and robotics~\cite{robot_self_explore_li, robot_self_explore_wang}.
These studies emphasize agents interacting with their environments to autonomously gather feedback and refine their capabilities. Such a learning paradigm aligns closely with reinforcement learning, where agents iteratively improve their performance through experience and reward-based feedback loops~\cite{RLagent24Feng, RLagent24zhang, agent_rl23Grigsby}. Curriculum learning are also effective learning strategies~\cite{Curriculum23Shi, voyager, Eureka23Ma}.
The integration of memory and self-learning mechanisms represents a critical step forward in enabling agents to operate effectively in real-world scenarios. This foundation sets the stage for developing more autonomous, intelligent systems capable of handling increasingly complex and varied tasks. Encouraged by these advancements, we aim to further explore the potential of self-learning agents in the domain of microservice management.

\noindent\textbf{LLM for Microservice Management.} Key operations in service management include data collection and anomaly detection~\cite{le2023log,jin2023assess, Yu2024monitor, Kang2024fault}, root cause analysis~\cite{ahmed2023recommending,chen2024automatic, Roy2024rca, Zhang2024rca, RCAgent24Wang}, and incident mitigation~\cite{qiao2023taskweaver,an2024nissist}.
Recent study have also explored the potential of fully autonomous, end-to-end service management systems. Though promising, existing LLMs still face challenge in addressing the complexity and variability of these systems~\cite{ACV24Zhang}. In this work, we take a first step by introducing an agent-based system designed to autonomously manage mciroservices from a self-learning perspective. Unlike prior approaches, our system operates without any initial knowledge of the microservices, progressively building and refining its service management capabilities over self-exploration with the service. This work aims to reduce reliance on specific domain knowledge while emphasizing adaptability and continuous learning. By presenting this work, we aim to inspire research interest within the community toward developing intelligent, self-learning agent systems capable of managing microservices autonomously.

\vspace{-3mm}
\section{Conclusion}
We present \ourmethod, a self-learning agent system for autonomic microservice management, reducing reliance on human input or static documentation. Through curriculum learning and iterative exploration, it progressively builds operational understanding and efficiently adapts to complex environments, as demonstrated with the Sock Shop prototype. Compared to manual efforts, \ourmethod offers a scalable alternative, significantly reducing the burden on human engineers for repetitive tasks like learning individual service setups, enabling engineers to focus on higher-value activities. While we have yet to experiment with more complex systems and challenges such as LLM limitations persist, future enhancements like efficient exploration task through parallelism and skill organization, hold the potential for broader applicability.

\bibliographystyle{ACM-Reference-Format}
\bibliography{sample-base}

\appendix

\newpage
\onecolumn  
\section{Prompts}

\subsection{Prompts for Our Framework}

The specific system prompts used for the task generation, and the memory mechanism are presented in Tables \ref{tab:prompt-task-generation}, and \ref{tab:prompt-memory}, respectively. The skills extracted via self-exploration in Trail 1 after 15 rounds is presented in Table \ref{tab:extracted-skills}, decoded for better human comprehension.

{\small
\begin{longtable}{p{\textwidth}}  
\caption{Prompt for task generation}\label{tab:prompt-task-generation}\\
\toprule  
\endfirsthead  
  
\hline  
\endhead  
  
\hline  
\multicolumn{1}{r}{\textit{Continued on next page}} \\ 
\endfoot  
  
\bottomrule  
\endlastfoot  

    - You are a Curriculum Builder who guides a LLM-powered management agent on how to gain knowledge/skills about managing a microservice through a series of exploration tasks. \\
    - The agent that you guide has some general knowledge about managing a microservice, but it is not familiar with the specific microservice system. \\
    - Your task is to design a series of tasks in peddagoical order, so that the agent can learn specific management knowledge/skills gradually. \\
    - You will have two inputs: the first one is the previous task list that you have generated for the agent to guide, and the second one is the learning areas that the agent to guide needs to learn. Refer to the `On the Input for Task List Generation` section for more details. \\
    - You need to output a new task list based on the previous trials. Follow the instructions in the `Instructions on Task List Generation` section to achieve this goal. \\
    - NOT generate any tasks on inspection or debugging on the components, as well as tools provided in the cluster. All of them are running well. \\ 
    \\
    \# Background Information about the Agent to Guide and the underlying microservice system \\
    - The agent to guide has the name \texttt{\`}\textcolor{blue}{\{agent\_name\}}\texttt{\`}. \\
    - The role of this agent is: managing the service component \texttt{\`}\textcolor{blue}{\{service\_component\_name\}}\texttt{\`}. \\
    - The microservice system is deployed on the Kubernetes cluster in the namespace \texttt{\`}\textcolor{blue}{\{namespace\_name\}}\texttt{\`}. \\
    \\
    \# On the Input for Task List Generation \\
    - The input contains two parts and is in the following JSON format: \\
    \begin{verbatim}
    "previous_task_list": [
      {
        "task_id": 1,
        "task_description": "(Caution: The following task is a self-learning exploration task which aims to help the 
        service management agents to gain knowledge and skills to manage the service component. 
        As the system is in a running state, you are supposed to obtain some informative response when fulfilling the task.) 
        Check the status of the ServiceX component",
        "trial_outcome": "SUCCESS"
      },
      {
        "task_id": 2,
        "task_description": "(Caution: The following task is a self-learning exploration task which aims to help the 
        service management agents to gain knowledge and skills to manage the service component. 
        As the system is in a running state, you are supposed to obtain some informative response when fulfilling the task.) 
        Check the latency of the ServiceX component",
        "trial_outcome": "FAILED: Lack of knowledge about ServiceX metrics collection via Prometheus"
      }
    ],
    "learning_areas": [
      "metric collection via Prometheus"
    ]
    \end{verbatim}
    - The `previous\_task\_list` contains the previous task list that you have generated for the agent to guide. Trial outcomes are also included in the input. The `trial\_outcome` field indicates whether the agent to guide has successfully completed the task or not. If the agent to guide has failed to complete a task, the reason for the failure might be provided in the `trial\_outcome` field. \\
    - The `learning\_areas` contains the knowledge/skills that the agent to guide needs to learn. \\
    - Note that both `previous\_task\_list` and `learning\_areas` are optional, namely, either or both of them can be empty. If both of them are empty, you need to generate a task list from scratch. \\ \\
    
    \# Instructions on Task List Generation \\
    
    \#\# General Instructions \\
    - Carefully review the `previous\_task\_list` and `learning\_areas` fields in the input. \\
    - If both of them are empty, you need to generate a task list from scratch. Refer to the `Instruction on Task List Generation from Scratch` subsection for more details. \\
    - If the `previous\_task\_list` is empty but contains `learning\_areas`, you need to generate a task list based on the learning areas. Refer to the `Instruction on Task List Generation based on Learning Areas` subsection for more details. \\
    - If the `previous\_task\_list` is not empty but does not contain any `learning\_areas`, you need to first decide learning areas based on the previous task list. Refer to the `Instruction on Task List Generation based on Previous Task List` subsection for more details. \\
    - If both `previous\_task\_list` and `learning\_areas` are not empty, you need to prioritize the `learning\_areas` over the potential learning areas derived from the previous task list. Refer to the `Instruction on Task List Generation based on Previous Task List and Learning Areas` subsection for more details. \\
    - When generating a list of tasks, you need to ensure that the tasks are pedagogically ordered, meaning that the tasks should be arranged in a way that allows the agent to guide to learn gradually. The tasks should also be related to the previous task list and the learning areas. If there are any conflicts between the previous task list and the learning areas, you need to resolve them by prioritizing the learning areas over the previous task list. \\
    - Each generated task should contain ONE specific task. Do NOT generate compound tasks like "Get both the CPU and memory usage of the component". Rather, generate two separate tasks: "Get the CPU usage of the component" and "Get the memory usage of the component". \\
    \#\# Instructions on Task List Generation from Scratch \\
    - Overall, the whole series of tasks should progress from those that do not change the state of the system to those that do. \\
    - For the tasks that do not change the state of the system, the goal is to help the agent to guide to get familiar with the service component and its basic metrics. You should generate them in the following order: \\
    - Check the basic information of the component, such as the version, status, and configuration. \\
    - Check the basic resource usage of the component, such as CPU and memory usage. They should be obtainable with built-in `kubectl` commands. \\
    - Find out what external tools are used in the cluster to monitor the component, such as Prometheus, Grafana, etc. \\
    - If Prometheus is used in the component, try to get familiar with the basic usage of Prometheus, such as how to query metrics related to latency and error rates. If the agent is struggling with metric collection via Prometheus, you should further break down the task into smaller steps, such as find out current Prometheus settings for this component, what the metrics are, and how to query them. \\

    \#\# Instruction on Task List Generation based on Learning Areas \\
    - Reason over the provided learning areas. If they are too vague, you need to expand them. \\
    - Map the learning areas to the tasks staged in the `Instruction on Task List Generation from Scratch` section. Then, refer to the `Instruction on Task List Generation from Scratch` section to generate a task list from that mapped staged task. For example, if the learning area is `how to get CPU utilization`, you can map it to the task of `Check the basic resource usage of the component`. Then, generate a task list from that task stage. \\
    
    \#\# Instruction on Task List Generation based on Previous Task List \\
    - Reason over the previous task list and decide the learning areas based on the previous task list. \\
    - Then, refer to the `Instruction on Task List Generation based on Learning Areas` section to generate a task list from that mapped staged task. \\
    
    \#\# Instruction on Task List Generation based on Previous Task List and Learning Areas \\
    - Reason over both the previous task list and the learning areas. If there are any conflicts between the previous task list and the learning areas, you need to resolve them by prioritizing the learning areas over the previous task list. \\
    - After resolving the conflicts, refer to the `Instruction on Task List Generation based on Learning Areas` section to generate a task list from that mapped staged task. \\ \\
    
    \# On the Generated Task List Format \\
    - The task list should be a JSON array, each element in the array is a JSON object. The JSON object should have the following fields: \\
    - `task\_id`: an integer, the ID of the task. \\
    - `task\_description`: a string, the description of the task. Add (Caution: The following task is a self-learning exploration task which aims to help the service management agents to gain knowledge and skills to manage the service component. As the system is in a running state, you are supposed to obtain some informative response when fulfilling the task.) before every task description. \\
    - Below is an example of a task list: \\
    \begin{verbatim}
    [
    {
      "task_id": 1,"
      "task_description": "(Caution: The following task is a self-learning exploration task which aims to help the
      service management agents to gain knowledge and skills to manage the service component. 
      As the system is in a running state, you are supposed to obtain some informative response when fulfilling the task.) 
      find the version of the component"
    },
    {
      "task_id": 2,"
      "task_description": "(Caution: The following task is a self-learning exploration task which aims to help the
      service management agents to gain knowledge and skills to manage the service component. 
      As the system is in a running state, you are supposed to obtain some informative response when fulfilling the task.) 
      find the CPU usage of the component"
    }
    ]
    \end{verbatim}
    
    \# Additional Cautions \\
    - Do NOT generate any tasks on inspection or debugging on the components, as well as tools provided in the cluster. All of them are running well. \\
    - Do NOT generate any explain or describe tasks. The tasks should be actionable and should require the agent to guide to perform some actions on the system. \\
\end{longtable}
}



{\small
\begin{longtable}{p{\textwidth}}  
\caption{Prompt for memory mechanism}\label{tab:prompt-memory}\\
\toprule  
\endfirsthead  
  
\hline  
\endhead  
  
\hline  
\multicolumn{1}{r}{\textit{Continued on next page}} \\ 
\endfoot  
  
\bottomrule  
\endlastfoot  
    - You are an intellient Experience/Skill Extractor, who extracts useful service management knowledge, experiences, and skills from a chat log between an agent-based management system and the managed service component in a Kubernetes environment. \\
    - The goal is to identify potentially useful experiences and skills for the agent-based management system so that it can leverage them in the future service management tasks. \\
    - You will be provided by two piece of input information: a chat log and a previous experience file. \\
    - Refer to the `Format of Chat Log` section to understand the format of the chat log. \\
    - Refer to the `Format of Extracted Experience` section to understand the format of the extracted experience. That would help to understand the provided previous experience file as well as the expected output format. \\
    - Refer to the `Instructions on Extracting Experience` section for the detailed instructions on how to extract the experiences and skills. \\ \\
    
    \# Format of Chat Log \\
    - The chat log provides the interaction history between the agent-based management system and the managed service component in a Kubernetes environment, with the following overall structure. Note that for convienence of referring to specific lines of the chat log, below we add line numbers to the chat log, which are not part of the original chat log. \\
    \begin{verbatim}
    [Line 1] Solving task...
    [Line 2] trigger (to <Service-X>):
    [Line 3] <Task Description>
    [Line 4] --------------------------------------------------------------------------------
    [Line 5] <Service-X> (to <Service-X>-group-manager):
    [Line 6] <Task Description>
    [Line 7] --------------------------------------------------------------------------------
    [Line 8] ==================== Step 1 ====================
    [Line 9] Next speaker: <Service-X>-assistant
    [Line 10] <Service-X>-assistant (to <Service-X>-group-manager):
    [Line 11] <Intermediate Analysis Results and Maintenance Codes to Solve the Task>
    [Line 12] --------------------------------------------------------------------------------
    [Line 13] Next speaker: <Service-X>-code-executor
    [Line 14] <Service-X>-code-executor (to <Service-X>-group-manager):
    [Line 15] <Code Excuation Results>
    [Line 16] --------------------------------------------------------------------------------
    [Line 17] ... (similar steps iterated multiple times)
    [Line 18] ==================== Step n ====================
    [Line 19] Next speaker: <Service-X>-assistant
    [Line 20] <Service-X>-assistant (to <Service-X>-group-manager):
    [Line 21] TERMINATE
    [Line 22] --------------------------------------------------------------------------------
    [Line 23] <Service-X> (to trigger):
    [Line 24] <Final Task Execution Summary and Response to the Task Trigger>
    \end{verbatim} \\
    - Line 1 - 3: Start of the chat log, indicating that the <Service-X> is triggered to solve a task with the description given in <Task Description>. \\
    - Line 5 - 22: The main part of the chat log, showing how the agent-based management system solves the task internally. \\
    - The agent-based management system consists of three components: <Service-X>-assistant, <Service-X>-code-executor, and <Service-X>-group-manager. \\
    - The <Service-X>-group-manager is mainly for receiving the task, passing information between the assistant and code-executor, and providing the final response to the task trigger. \\
    - The <Service-X>-assistant and <Service-X>-code-executor are the key components interacting to solve the task in each step. \\
    - Each step starts with the <Service-X>-assistant providing an analysis or maintenance codes to solve the task, followed by the <Service-X>-code-executor executing the codes and providing the results. \\
    - The whole task can take a series of steps, from Step 1 to Step n, until the task is terminated with the keyword "TERMINATE". \\
    - Line 23 - 24: End of the chat log, showing the final summary of the task execution and the response to the task trigger.\\ \\
    
    \# Format of Extracted Experience \\
    - The extracted experience file has the following format: \\
    \# Experience about Monitoring Kubernetes Components \\
    \#\# Command \\
    - Command 1: `<command 1>`, Explanation for when and how to use this command 1. \\
    - Command 2: `<command 2>`, Explanation for when and how to use this command 2. \\
    \#\# Reflection \\
    - Reflection 1: <Summary of the reflection on past failures or successes>. \\
    - Reflection 2: <Summary of the reflection on past failures or successes>. \\
    \#\# Configuration \\
    - Configuration 1: <list one key configuration of the managed service component>. \\
    - Configuration 2: <list one key configuration of the managed service component>. \\
    \#\# Conflicted Experience Requiring Resolution \\
    - Conflict 1: <description of the conflict and suggested resolution>. \\
    - Conflict 2: <description of the conflict and suggested resolution>. \\
    - The `Command` section lists the successful commands related to the previously solved task, with a brief explanation of when and how to use each command. \\
    - The `Reflection` section provides a summary of the reflection on past failures or successes, which helps to understand the key points learned from the past task. \\
    - The `Configuration` section lists the key configurations of the managed service component that are essential for the task execution. Plz not use etc. to replace the actual configuration, you must list them all. \\\
    - The `Conflicted Experience Requiring Resolution` section lists the conflicting experiences that need to be resolved before using them in future tasks. \\ \\
    \# Instructions on Extracting Experience \\
    - First, carefully read through the chat log by understanding the task description, the interactions between the components, and the final task execution summary. \
    - Then, identify the following types of information to extract: \\
    - Commands: Successful commands related to the current solving task. \\
      - A successful command comes from the assistant and executed by the code-executor. The result from the code-executor should be successful: Not empty, No error message, the output is expected as the assistant scheduled. \\
      - If empty (like empty list[] without actual data even with success code, string ""), error message, or unexpected output, the command should not be extracted. \\
    - Reflections: Key information or reasons for the failure of some commands or tasks. \\
      - If the some commands are not successful, you should provide the possible reason for the failure. \\
      - After some commands is executed, the chat log get stuck, no further commands are executed, which means this kinds of commands will not be successful and must never be executed again, you should provide insturctions on not executing this kind of commands. \\
    - Configurations: Key configuration information for the system, tools, etc. \\
     - Code execution results often contain lots of useful service configuration information, which can be extracted as key configurations. \\
     - When extracting key configurations from the code execution results, you need to be specific, concise and organized. Avoid outputing too general and vague information. \\
    - Before outputing the updated experience, you should consolidate the newly-extracted experience with the previous experience file with the following considerations: \\
    - The newly-extracted experience should be clear, concise, and relevant to the task described in the chat log. \\
    - If the newly-extracted experience is similar to the previous experience, you can combine them together. \\
    - If the newly-extracted experience is not present in the previous experience, you should add it as a new experience and classify it to the corresponding section (Command, Reflection, or Configuration). \\
    - If the newly-extracted experience is conflicting with the previous experience, you should add one item in the `Conflicted Experience Requiring Resolution` section by describing the conflict and suggesting a resolution. \\
    - If the newly-extracted experience is helpful for understanding the previous experience conflict, you can add it to the `Reflection` section and remove the corresponding conflicted item in the `Conflicted Experience Requiring Resolution` section. Note that if one task can be solved with several methods, those working methods are not considerred as conflicting experiences. \\
    - When outputing the final updated experience, they should be formatted according to the `Format of Extracted Experience` section. \\ \\
    
    \# Example (Not add Example content into your output, only reference for understanding) \\
    \begin{verbatim}
    <Input>  
    The experience source is from 2025-01-08.
    
    The chat log is as follows:
    
    '''
    Solving task...
    trigger (to Service-X):
    find the current image version of the service
    --------------------------------------------------------------------------------
    Service-X (to Service-X-group-manager):
    find the current image version of the service
    --------------------------------------------------------------------------------
    ==================== Step 1 ====================
    Next speaker: Service-X-assistant
    Service-X-assistant (to Service-X-group-manager):
    here is the command to find the current image version of the service: 
    ```bash
    kubectl get pods -n service-namespace -o jsonpath="{.items[*].spec.containers[*].image}"
    ```
    --------------------------------------------------------------------------------
    Next speaker: Service-X-code-executor
    Service-X-code-executor (to Service-X-group-manager):
    nginx:1.19.1
    --------------------------------------------------------------------------------
    ==================== Step 2 ====================
    Next speaker: Service-X-assistant
    Service-X-assistant (to Service-X-group-manager):
    TERMINATE
    --------------------------------------------------------------------------------
    Service-X (to trigger):
    the current image version of the service is nginx:1.19.1
    '''
    
    Here is the previous experience you have extracted:
    
    '''
    \end{verbatim}
    \# Experience about Monitoring Kubernetes Components \\
    \#\# Command \\
    - Command 1: `kubectl get namespaces`, Used for listing all namespaces present in the Kubernetes cluster to identify where services are deployed. \\
    \#\# Reflection \\
    - Reflection 1: Always check the namespace before executing any command to avoid unintended consequences.
    \#\# Configuration \\
    - Configuration 1: <Service-X> is under the namespace `service-namespace`. \\
    \#\# Conflicted Experience Requiring Resolution \\
    - Conflict 1: Previously, sometimes we can use `kubectl get pods` to find the current image version of the service, but it may not work for all services. Need to find a more reliable way to get the image version. \\ \\
    
    <Output> \\
    \# Experience about Monitoring Kubernetes Components \\
    \#\# Command \\
    - Command 1: `kubectl get namespaces`, Used for listing all namespaces present in the Kubernetes cluster to identify where services are deployed. \\
    - Command 2: `kubectl get pods -n service-namespace -o jsonpath="{.items[*].spec.containers[*].image}"`, Used for finding the current image version of the service. \\
    \#\# Reflection \\
    - Reflection 1: Always check the namespace before executing any command to avoid unintended consequences. \\
    - Reflection 2: The command `kubectl get pods -n service-namespace -o jsonpath="{.items[*].spec.containers[*].image}"` is a reliable way to find the current image version of the service. \\
    \#\# Configuration \\
    - Configuration 1: <Service-X> is under the namespace `service-namespace`. \\
    - Configuration 2: The current image version of the service is `nginx:1.19.1`. \\
    \#\# Conflicted Experience Requiring Resolution \\
    - None \\
\end{longtable}
}



{\small
\begin{longtable}{p{\textwidth}}  
\caption{Extracted Skills}\label{tab:extracted-skills}\\
\toprule  
\endfirsthead  
  
\hline  
\endhead  
  
\hline  
\multicolumn{1}{r}{\textit{Continued on next page}} \\ 
\endfoot  
  
\bottomrule  
\endlastfoot  
    \# Experience about Monitoring Kubernetes Components \\
    \#\# Command \\
    - Command 1: kubectl get deployments --all-namespaces | grep catalogueUsed for quickly locating the “catalogue” deployment across all namespaces in the Kubernetes cluster. \\
    - Command 2: kubectl describe deployment catalogue -n sock-shop | grep ImageUsed for retrieving the container image version of the “catalogue” deployment within the “sock-shop” namespace. \\
    - Command 3: kubectl get pods --all-namespaces | grep catalogueUsed for listing pods named or labeled “catalogue” across all namespaces in the Kubernetes cluster. \\
    - Command 4: kubectl describe pod catalogue-5b877d88b4-g9tc4 -n sock-shopUsed for describing the details (status, container image, resource usage, probes) of the specific “catalogue” pod. \\
    - Command 5: report\_result(component='catalogue', message='...', message\_type='RESPONSE')Used within a Python script to send the final status or version information of the “catalogue” component back to the manager. \\
    - Command 6: kubectl get pods -n sock-shop | grep catalogueUsed for listing the “catalogue” pods within the “sock-shop” namespace. \\
    - Command 7: kubectl top pod catalogue-5b877d88b4-g9tc4 -n sock-shopUsed for checking the current CPU and memory usage of the “catalogue” pod, provided that Metrics Server is installed in the cluster. \\
    - Command 8: curl 'http://192.168.58.2:31090/api/v1/label/\_\_name\_\_/values'Used for listing all metric names available in the Prometheus server, helping identify which metrics are relevant for a given service. \\
    - Command 9: curl 'http://192.168.58.2:31090/api/v1/label/job/values'Used for listing all possible values of the “job” label from Prometheus, so one can discover valid job names such as “sock\mbox{-}shop/catalogue.” \\
    - Command 10: curl 'http://192.168.58.2:31090/api/v1/query?query=http\_requests\_total\{job="sock\mbox{-}shop/catalogue"\}'Used for querying the “http\_requests\_total” metric from Prometheus for the “catalogue” job. Even if successful, it may return empty data if there is no current traffic. \\
    - Command 11: curl 'http://192.168.58.2:31090/api/v1/query?query=rate(http\_request\_duration\_seconds\_sum\{job="sock\mbox{-}shop/catalogue"\}[5m]) / rate(http\_request\_duration\_seconds\_count\{job="sock-shop/catalogue"\}[5m])'Used for retrieving the average response time for the “catalogue” service from Prometheus over the last five-minute window. This command can return empty results if there is no traffic to the service. \\
    - Command 12: curl 'http://192.168.58.2:31090/api/v1/query?query=rate(http\_requests\_total\{status=\~"4..",job="sock\mbox{-}shop/catalogue"\}[5m])'Used for querying Prometheus to retrieve the 4xx error rate for the “catalogue” service. If no 4xx errors are present, it returns an empty “result” array. \\
    - Command 13: curl 'http://192.168.58.2:31090/api/v1/query?query=rate(http\_requests\_total\{job="sock\mbox{-}shop/catalogue",status=\~"5.."\}[5m])'\\Used for querying Prometheus to retrieve the 5xx error rate for the “catalogue” service. If no 5xx errors are present, it returns an empty “result” array. \\
    - Command 14: curl 'http://192.168.58.2:31090/api/v1/query?query=histogram\_quantile(0.95, sum(rate(request\_duration\_seconds\_bucket\\ \{job="sock\mbox{-}shop/catalogue"\}[5m])) by (le))'Used for retrieving the 95th percentile (p95) latency for the “catalogue” service from Prometheus. This command calculates the p95 latency by combining histogram\_quantile(0.95, ...) with sum(rate(...)) of the histogram buckets over a five-minute window. \\
    - Command 15: curl 'http://192.168.58.2:31090/api/v1/query?query=process\_cpu\_seconds\_total\{job="sock\mbox{-}shop/catalogue"\}'Used for querying the total CPU usage from Prometheus for the “catalogue” service. This command retrieves the “process\_cpu\_seconds\_total” metric by properly URL-encoding the job label to avoid parse errors. \\
    - Command 16: curl 'http://192.168.58.2:31090/api/v1/query?query=process\_resident\_memory\_bytes\{job="sock\mbox{-}shop/catalogue"\}'Used for querying the memory usage from Prometheus for the “catalogue” service by retrieving the “process\_resident\_memory\_bytes” metric with proper URL encoding. \\
    - Command 17: curl 'http://192.168.58.2:31090/api/v1/query?query=nodejs\_active\_requests\_total\{job="sock\mbox{-}shop/catalogue"\}'Used for retrieving the number of active requests (connections) for the “catalogue” service from Prometheus. This command may return an empty result if there are no current active requests. \\ \\
    \# Reflection \\
    - Reflection 1: All commands executed successfully without errors. Filtering deployments or pods by name with “grep” is quick and worked well here, but one should ensure that the target component name is sufficiently unique to avoid accidental matches. \\
    - Reflection 2: The command kubectl top pod  -n  can be used to view real-time CPU and memory usage of a pod when Metrics Server is available in the cluster. Recent usage retrieval showed 9Mi of memory usage for the “catalogue” component, confirming that the approach works as expected. \\
    - Reflection 3: Multiple Prometheus curl queries failed with parse errors (e.g., invalid parameter “query”) due to unencoded special characters in the URL. Encoding square brackets and braces (for example, \%5B and \%5D) is crucial to constructing valid Prometheus queries. \\
    - Reflection 4: The correct Prometheus “job” label for the catalogue service is sock-shop/catalogue. However, querying for recent request throughput returned empty results, possibly indicating no traffic to the “catalogue” component in that time span. \\
    - Reflection 5: Querying the average response time (rate(http\_request\_duration\_seconds\_sum)/rate(http\_request\_duration\_seconds\_count)) for the “catalogue” service also returned an empty result. This further suggests no current traffic or metrics for the “catalogue” component at this time. \\
    - Reflection 6: Additional queries for the 4xx and 5xx error rate using rate(http\_requests\_total\{status="4.."\}[5m]) \\ and rate(http\_requests\_total\{status="5.."\}[5m]) returned empty data, indicating no recent error codes. Proper URL encoding remains essential to avoid parse errors. \\
    - Reflection 7: Successfully used histogram\_quantile(0.95, sum(rate(request\_duration\_seconds\_bucket{job="sock-shop/catalogue"}[5m])) by (le))) to retrieve the p95 latency for the “catalogue” service—approximately 4.82 milliseconds. Ensuring the correct “job” label and valid histogram metric is crucial for accurate percentile calculations. \\
    - Reflection 8: “process\_resident\_memory\_bytes” is a relevant Prometheus metric for retrieving memory usage. A query with unencoded braces initially failed, but proper URL encoding fixed the issue and returned the expected memory usage value for the “catalogue” service. \\
    - Reflection 9: The “nodejs\_active\_requests\_total” metric returned an empty result for the “catalogue” service, indicating there were no active connections at that moment. This is not necessarily an error; the service may simply have had no traffic at the time of the query. \\ \\

    \# Configuration \\
    - Configuration 1: Container image of the “catalogue” component is weaveworksdemos/catalogue:0.3.5. \\
    - Configuration 2: The container requests 100m CPU and 100Mi memory, while the limits are 200m CPU and 200Mi memory. \\
    - Configuration 3: Both liveness and readiness probes are configured with an HTTP GET on /health, a 10s initial delay, 1s timeout, 3s period, requiring 1 success and allowing 3 failures. \\
    - Configuration 4: The container command is /app with the argument -port=80. \\

    \# Conflicted Experience Requiring Resolution \\
    - None \\
\end{longtable}
}

\end{document}